\documentclass[apss,preprint,groupedaddress,superscriptaddress,showpacs,amsmath,amssymb,amsfonts,floatfix,endfloats]{aastex}
\usepackage{graphicx}
\usepackage{dcolumn}
\usepackage{bm}
\usepackage{amsmath}
\usepackage{amssymb}
\usepackage{bbm}
\usepackage{mathrsfs}
\usepackage{epsfig}
\usepackage{amsmath,amsfonts,amssymb}
\usepackage{fixltx2e}
\usepackage[mathlines]{lineno}

\shorttitle{Collapsed Cores in Globular Clusters}
\shortauthors{Djorgovski et al.}

\begin{document}


\title{Kinetic theory of acoustic-like modes in nonextensive pair plasmas}


\author{E. Saberian\altaffilmark{1}}
\email{s.esaberian@azaruniv.edu}
\affil{Department of Physics, Faculty of Basic Sciences, University of Neyshabur, P.O.Box: 91136-599, Neyshabur, Iran}

\and

\author{A. Esfandyari-Kalejahi}
\affil{Department of Physics, Faculty of Basic Sciences, Azarbaijan Shahid Madani University, P.O.Box: 53714-161, Tabriz, Iran}
\email{ra-esfandyari@azaruniv.edu}


\altaffiltext{1}{Visiting Researcher in Department of Physics, Faculty of Basic Sciences, Azarbaijan Shahid Madani University, P.O.Box: 53714-161, Tabriz, Iran}


\date{\today}

\begin{abstract}
The low-frequency acoustic-like modes in a pair plasma (electron-positron or pair-ion) is studied by employing a kinetic theory model based on the Vlasov and Poisson's equation with emphasizing the Tsallis's nonextensive statistics. The possibility of the acoustic-like modes and their properties in both fully symmetric and temperature-asymmetric cases are examined by studying the dispersion relation, Landau damping and instability of modes.
The resultant dispersion relation in this study is compatible with the acoustic branch of the experimental data [W. Oohara, D. Date, and R. Hatakeyama, Phys. Rev. Lett. 95, 175003 (2005)], in which the electrostatic waves have been examined in a pure pair-ion plasma.
Particularly, our study reveals that the occurrence of growing or damped acoustic-like modes depends strongly on the nonextensivity of the system as a measure for describing the long-range Coulombic interactions and correlations in the plasma. The mechanism that leads to the unstable modes lies in the heart of the nonextensive formalism yet, the mechanism of damping is the same developed by Landau. Furthermore, the solutions of acoustic-like waves in an equilibrium Maxwellian pair plasma are recovered in the extensive limit ($q\rightarrow1$), where the acoustic modes have only the Landau damping and no growth.
\end{abstract}


\keywords{Pair plasmas: Kinetic theory of plasma waves: Waves, oscillations, and instabilities in plasmas}



\section{Introduction}

Studying the pair plasmas has been an important challenge for many plasma physicists in two past decades.
As we know, the difference between the electron and ion masses in an ordinary electron-ion plasma (in general, multi-component plasma with both light and heavy particles) gives rise to different time-space scales which are used to simplify the analysis of low- and high-frequency modes. Such time-space parity disappears when studying a pure pair plasma which consisting of only positive- and negative-charged particles with an equal mass, because the mobility of the particles in the electromagnetic fields is the same.
Pair plasmas consisting of electrons and positrons have attracted an especial of interest because of their significant applications in astrophysics.
In fact, electron-positron plasmas play an important role in the physics of a number of astrophysical situations such as active galactic nuclei \citep{121,122}, pulsar and neutron star magnetosphere \citep{31,32,33}, solar atmosphere \citep {4}, accretion disk \citep{5}, black holes \citep{6}, the early universe \citep{781,782} and many others. For example, the detection of circularly polarized radio emission from the jets of the archtypal quasar 3C297, indicates that electron-positron pairs are an important component of the jet plasma \citep{9}. Similar detections in other radio sources suggest that, in general, extragalactic radio jets are composed mainly of an electron-positron plasma \citep{9}. Furthermore, it has been suggested that the creation of electron-positron  plasma in pulsars is essentially by energetic collisions between particles which are accelerated as a result of electric and magnetic fields in such systems \citep{1011121,33,1011123}. On the other hand, the successful achievements for creation of the electron-positron plasmas in laboratories have been frequently reported in the scientific literatures \citep{experiments1,experiments2,experiments3,experiments4,
experiments5,experiments6,experiments7}.
In this regard, many authors have concentrated on the relativistic electron-positron plasmas \citep{221,222,223,224,225,226} because of its occurrence in astrophysics and encountering with positron as an antimatter in high-energy physics. However, there are many experiments that confirm the possibility of nonrelativistic electron-positron plasmas in laboratory \citep{231,232}.
It has been observed that the annihilation time of electron-positron pairs in typical experiments is often long compared with typical confinement times \citep{1920}, showing that the lifetime of electron-positron pairs in the plasma is much longer than the characteristic time scales of typical oscillations. The long lifetime of electron-positron pairs against pair annihilation indicates that many collective modes can occur and propagate in an electron-positron plasma.

Although pair plasmas consisting of electrons and positrons have been experimentally produced, however, because of fast annihilation and the formation of positronium atoms and also low densities in typical electron-positron experiments, the identification of collective modes in such experiments is practically very difficult.
To resolve this problem, one may experimentally deal with a pure pair-ion plasma instead of a pure electron-positron plasma for identification of the collective modes. An appropriate experimental method has been developed by Oohara and Hatakeyama \citep{fullerene} for the generation of pure pair-ion plasmas consisting only positive and negative ions with equal masses by using fullerenes $\mathrm{C}_{60}^{-}$ and $\mathrm{C}_{60}^{+}$. The fullerenes are molecules containing 60 carbon atoms in a very regular geometric arrangement, and so a fullerene pair plasma is physically akin to the an electron-positron plasma, without having to worry about fast annihilation.
By drastically improving the pair-ion plasma source in order to excite effectively the collective modes, Oohara \emph{et al.} \citep{Oohara} have experimentally examined the electrostatic modes propagating along the magnetic-field lines in a fullerene pair plasma.

In exploring the electrostatic modes in a pair plasma, most of authors have merely studied the high frequency Langmuir-type oscillation in a pure electron-positron plasma \citep{Tsytovich,Iwamato,Zank,Verheest} or in a pure pair-ion plasma \citep{Vranjes} via the theoretical studies.
Particularly, Iwamato \citep{Iwamato} and Veranjes and Poedts \citep{Vranjes} have studied the longitudinal modes in a pair plasma only in the case in which the phase velocity of the wave is much larger than the thermal velocity of the particles which leads to the Langmuir-type waves.
However, in the experiment of Oohara \emph{et al.} \citep{Oohara}, three kinds of electrostatic modes have been observed from the obtained dispersion curves: a relatively low-frequency band with nearly constant group velocity (the acoustic waves), an intermediate-frequency backward-like mode (to our knowledge, with the lack of a satisfactory theoretical explanation), and the Langmuir-type waves in a relatively high-frequency band. This experiment indicates that in a pure pair plasma, besides of the Langmuir-type waves, the acoustic-like modes are possible in practice.
There, Oohara \emph{et al.} \citep{Oohara} have briefly discussed some aspects of their experimental results by using a theoretical two-fluid model.
Here, our goal is to investigate the possibility and the properties of the intriguing acoustic-like modes in a pair plasma (in both symmetric and asymmetric cases) by using a kinetic theory model and to argue some properties of these modes in a subtler manner.
It is to be noted that the only asymmetry in a pure pair plasma may arise from a difference in temperatures of species.
Physically, the temperature-asymmetry in a pair plasma may arise from the typical experimental procedure in which a pair plasma is produced in the laboratory. For example, an effective technique for creating an electron-positron plasma in the laboratory is as follows: at first, we may obtain a positron plasma through scattering from a buffer gas into a panning trap \citep{positron plasma1,positron plasma2}; using this technique, the positrons can be stored at densities of order $10^{7}cm^{-3}$ and lifetime of order $10^{3}sec$ in the recent experiments. Then, an electron-positron plasma with sufficient stability can be produced by injecting a low-energy electron beam into the positrons \citep{ep production1,ep production2,ep production3,ep production4}.
For our purpose, we assume that the phase speed of the acoustic-like modes lies in the vicinity of the thermal velocities of the species (in fact, between the thermal velocities of the two species). The situation is somewhat similar to the case in which the possibility and properties of the electron-acoustic waves in a two-temperature (cold and hot) electron plasma is examined \citep{electron acoustic1,electron acoustic2,electron acoustic3,electron acoustic4,electron acoustic5}.

It is often observed that the physical distribution of particles in space plasmas as well as in laboratory plasmas are not exactly Maxwellian and particles show deviations from the thermal distribution \citep{non-Maxwellian1,non-Maxwellian2}. Presence of nonthermal particles in space plasmas has been widely confirmed by many spacecraft measurements \citep{super1,super2,super3,super4}. In many cases, the velocity distributions show non-Maxwellian tails decreasing as a power-law distribution in particle speed.
Several models for phase space plasma distributions with superthermal wings or other deviations from purely Maxwellian behavior have become rather popular in recent years, like the so-called kappa ($\kappa$) distribution which was introduced initially by Vasyliunas in 1968 \citep{Vasyliunas} for describing plasmas out of the thermal equilibrium such as the magnetosphere environments and the Solar winds \citep{Maksimovic}, or the nonthermal model advanced by Cairns \emph{et al.} in 1995 \citep{Cairns} which was introduced at first for an explanation of the solitary electrostatic structures involving density depletions that have been observed in the upper ionosphere in the auroral zone by the Freja satellite \citep{Freja}, and also the nonextensive model which go under the name of Tsallis. In the following we want to briefly review the formalism of the Tsallis model and to argue why it is preferred, rather than that of the Cairns and kappa model.

From a statistical point of view, there are numerous studies indicating the breakdown of the Boltzmann-Gibbs (BG) statistics for description of many systems with long-range interactions, long-time memories and fractal space-time structures (see, e.g., \citet{breakdown1,breakdown2,breakdown3,breakdown4}).
Generally, the standard BG extensive thermo-statistics constitutes a powerful tool when microscopic interactions and memories are short ranged and the environment is an Euclidean space-time, a continuous and differentiable manifold.
Basically, systems subject to the long-range interactions and correlations and long-time memories are related to the nonextensive statistics where the standard BG statistics and its Maxwellian distribution do not apply.
The plasma environments in the astrophysical and laboratorial systems are obviously subject to spatial and temporal long-range interactions evolving in a non-Euclidean space-time that make their behavior nonextensive. A suitable generalization of the Boltzmann-Gibbs-Shannon (BGS) entropy for statistical equilibrium was first proposed by Reyni \citep{Reyni} and subsequently by Tsallis \citep{Tsallis1,Tsallis2}, preserving the usual properties of positivity, equiprobability and irreversibility, but suitably extending the standard extensivity or additivity of the entropy to nonextensivity.

The nonextensive generalization of the BGS entropy which proposed by Tsallis in 1988 \citep{Tsallis1,Tsallis2} is given by the following expression:
\begin{equation}
S_{q}=k_{B}\frac{1-\sum_{i}  p_{i}^{q}}{q-1},
\end{equation}
where $k_{B}$ is the standard Boltzmann constant, $\{p_{i}\}$ denotes the probabilities of the microstate configurations and $q$ is a real parameter quantifying the degree of nonextensivity. The most distinctive feature of $S_{q}$ is its pseudoadditivity. Given a composite system $A+B$, constituted by two subsystems $A$ and $B$, which are independent in the sense of factorizability of the joint microstate probabilities, the Tsallis entropy of the composite system $A+B$ satisfies $ S_{q}(A+B)= S_{q}(A)+S_{q}(B)+(1-q)S_{q}(A)S_{q}(B) $. In the limit of $ q\rightarrow1$, $S_{q}$ reduces to the celebrated logarithmic Boltzmann-Gibbs entropy $S=-k_{B} \sum_{i} p_{i}\ln p_{i}$, and the usual additivity of entropy is recovered. Hence, $|1-q|$ is a measure of the lack of extensivity of the system.
There are numerous evidences exhibiting that the nonextensive statistics, arising from $S_{q}$, is a better framework for describing many physical systems such as the galaxy clusters \citep{galaxy cluster}, the plasmas \citep{evidence for plasmas1,evidence for plasmas2}, the turbulent systems \citep{turbulence1,turbulence2,turbulence3}, and so on, in which the system shows a nonextensive behavior as a result of long-range interactions and correlations. The experimental results in such systems display a non-Maxwellian velocity distribution for the particles \citep{non-Maxwellian1,non-Maxwellian2}. The functional form of the velocity distribution in the Tsallis formalism may be derived through a nonextensive generalization of the Maxwell ansatz \citep{Maxwell ansatz}, or through the maximizing Tsallis' entropy under the constraints imposed by normalization and the energy mean value \citep{mean value constraint1,mean value constraint2}. Furthermore, from a nonextensive generalization of the ``molecular chaos hypothesis'', it is shown that the equilibrium $q$-nonextensive distribution is a natural consequence of the \emph{H} theorem \citep{H-theorem}.

It is to be noted that the empirically derived kappa distribution function in space plasmas is equivalent to the $q$-distribution function in Tsallis nonextensive formalism, in the sense that the spectrum of the velocity distribution function in both models show the similar behavior and, in fact, both the kappa distribution and the Tsallis $q$-nonextensive distribution describe deviations from the thermal distribution. Particularly, Leubner in 2002 \citep{Leubner} showed that the distributions very close to the kappa distributions are a consequence of the generalized entropy favored by the nonextensive statistics, and proposed a link between the Tsallis nonextensive formalism and the kappa distribution functions. In fact, relating the parameter $q$ to $\kappa$ by formal transformation $\kappa=1/(1-q)$ \citep{Leubner} provides the missing link between the $q$-nonextensive distribution and the $\kappa$-distribution function favored in space plasma physics, leading to a required theoretical justification for the use of $\kappa$-distributions from fundamental physics.
Furthermore, Livadiotis and McComas in 2009 \citep{Livadiotis} examined how kappa distributions arise naturally from the Tsallis statistical mechanics.
On the other hand, the nonthermal distribution function introduced by Cairns et al. \citep{Cairns} is a proposal function to model an electron distribution with a population of energetic particles. It is especially appropriate for describing the nonlinear propagation of large amplitude electrostatic excitations such as solitary waves and double layers which are very common in the magnetosphere. However, the lack of a statistical foundation behind this proposal function is clearly seen, leading to less attention to it rather than the kappa function and the Tsallis distribution. Anyway, the $q$-nonextensive formalism, with a powerful thermo-statistics foundation and numerous experimental evidences, may cover many features of the other nonthermal models and provide a good justification for its preference over the other models. It has considerably extended both statistical mechanics formalism and its range of applicability. The interested reader may refer to the Refs. \citep{why Tsallis1,why Tsallis2,why Tsallis3,why Tsallis4} where the significance, historical background, physical motivations, foundations and applications of the nonextensive thermo-statistics have been discussed in detail.

The problem of waves, Landau damping and instabilities in typical plasmas have been investigated by some authors in the framework of the Tsallis nonextensive statistics \citep{waves-Tsallis1,waves-Tsallis2,waves-Tsallis3,waves-Tsallis4,Saberian}.
Particularly, it is to be noted that the physical state described by the $q$-nonextensive distribution in the Tsallis's statistics is not exactly the thermodynamic equilibrium \citep{waves-Tsallis4}. In fact, the deviation of $q$ from unity quantifies the degree of inhomogeneity of the temperature $T$ via the formula $k_{B}\nabla T+(1-q)Q_{\alpha}\nabla \phi=0 $ \citep{inhomogeneity of T}, where $Q_{\alpha}$ denotes the electric charge of specie $\alpha$, and $\phi$ is the electrostatic potential. In other words, the nonextensive statistics describes a system that have been evolved from a nonequilibrium stationary state with inhomogeneous temperature which contains a number of nonthermal particles.

In the present work, we attempt to investigate the possibility of the acoustic-like modes in a field-free and collisionless pair plasma (electron-positron or pair-ion) and to discuss the damping and instability of modes in the context of the Tsallis' nonextensive statistics. In Sec. 2, a kinetic theory model based on the linearized Vlasov and Poisson's equations is applied for deriving the dielectric function ($D(k,\omega)$) for longitudinal waves in an unmagnetized pair plasma. We then find the solutions of $D(k,\omega)=0$ for the acoustic-like waves with the constraint of weak damping or growth by considering a $q$-nonextensive distribution for stationary state of the plasma, as demonstrated in Sec. 3. The dispersion relation, Landau damping and instability of the acoustic-like modes are discussed in Sec. 4. Finally, a summary of our results is given in Sec. 5.

\section{The model equations}
In this section, we present a brief review of kinetic equations for describing the electrostatic collective modes specialized to a pair plasma (electron-positron or pair-ion) with the constraint of weak damping or growth.

We consider a spatially uniform field-free pair plasma at the equilibrium state. If at a given time $t=0$ a small amount of charge is displaced in the plasma, the initial perturbation may be described by

\[f_{\alpha}(t=0)=f_{0,\alpha}(\vec{v})+f_{1,\alpha}(\vec{x},\vec{v},t=0),\ \  \ \ f_{1,\alpha}\ll f_{0,\alpha},\]
where $f_{0,\alpha}$ corresponds to the unperturbed and time-independent stationary distribution and $f_{1,\alpha}$ is the corresponding perturbation about the equilibrium state. Here, $\alpha$ stands for electrons and positrons ($\alpha=e^{\pm}$) or fullerene pairs ($\alpha=\mathrm{C}_{60}^{\pm}$). We assume that the perturbation is electrostatic and the displacement of charge gives rise to a perturbed electric but no magnetic field. With this assumption, the time development of $f_{1,\alpha}(\vec{x},\vec{v},t)$ is given by the solution of the linearized Vlasov and Poisson's equations as follows \citep{Landau paper,Krall}:
\begin{equation}
\frac{\partial f_{1,-}}{\partial t}+\vec{v}\cdot\frac{\partial f_{1,-}}{\partial \vec{x}}+\frac{e}{m}\nabla\phi_{1}\cdot\frac{\partial f_{0,-}}{\partial \vec{v}}\;=\;0,
\label{f1-}
\end{equation}
\begin{equation}
\frac{\partial f_{1,+}}{\partial t}+\vec{v}\cdot\frac{\partial f_{1,+}}{\partial \vec{x}}-\frac{e}{m}\nabla\phi_{1}\cdot\frac{\partial f_{0,+}}{\partial \vec{v}}\;=\;0,
\label{f1+}
\end{equation}
\begin{equation}
\nabla^{2}\phi_{1}=4\pi n_{0}e\int (f_{1,-}-f_{1,+})\,\mathrm{d}\vec{v},
\label{Poisson}
\end{equation}
where $e$, $m$ and $n$ denote, respectively, the absolute charge, mass and number density of the pairs and $\phi_{1}$ is the electrostatic potential produced by the perturbation. Here, we have labeled the distribution function of negative and positive pairs with the subscripts $\pm$. This set of linearized equations for perturbed quantities may be solved simultaneously to investigate the plasma properties for the time intervals shorter than the binary collision times. Specially, we can study the properties of the plasma waves whose oscillations period are much less than a binary collision time. The standard technique for simultaneously solving the differential equations (2)-(4) is the method of integral transforms, as developed for the first time by Landau in the case of an ordinary electron-ion plasma \citep{Landau paper,Krall}.
Another simplified method of solving the Vlasov-Poisson's equations for the longitudinal waves, with the frequency $\omega$ and the wave vector $\vec{k}$, is to assume that the solution has the form
\begin{equation}
\begin{array}{l}
 f_{1,\alpha}(\vec{x},\vec{v},t)=f_{1,\alpha}(\vec{v}) e^{i(\vec{k}\cdot \vec{x}-\omega t)},   \ \ \  \alpha=e^{\pm} \ \ \mathrm{or} \ \ \mathrm{C}_{60}^{\pm} , \\
\phi_{1}(\vec{x},t)=\phi_{1} e^{i(\vec{k}\cdot \vec{x}-\omega t)}.
\end{array}
\label{simplified}
\end{equation}

Without loss of the generality, we consider the $x$-axis to be along the direction of the wave vector $\vec{k}$, and let $v_{x}=u$.
Then, by applying the Eq. (\ref{simplified}) and solving the Eqs. (\ref{f1-})-(\ref{Poisson}) we find the dispersion relation for longitudinal waves in a pair plasma as follows
\begin{equation}
D(k,\omega)=1-\frac{4 \pi n_{0} e^2}{mk^2} \int \frac{\frac{\partial}{\partial u}(f_{0,-} (u)+f_{0,+}(u))}{u-\frac{\omega}{k}}\,\mathrm{d}u=0, \label{DR}
\end{equation}
where $D(k,\omega)$ is the dielectric function of a field-free pair plasma for the longitudinal oscillations. We then can investigate the response of the pair plasma to an arbitrary perturbation via the response dielectric function $D(k,\omega)$.
In general, the frequency $\omega$ which satisfies the dispersion relation $D(k,\omega)=0$ is complex, i.e., $\omega=\omega_{r}+i\omega_{i}$. However, in many cases $Re[\omega(k)]\gg Im[\omega(k)]$, and the plasma responds to the perturbation a long time after the initial disturbance with oscillations at a range of the well-defined frequencies. These are the normal modes of the plasma, in the sense that they are the nontransient response of the plasma to an initial perturbation. We can determine the normal modes of the plasma via the dispersion relation $D[k,\omega(k)]=0$, which gives the frequency of the plasma waves as a function of the wave number $k$ or vice versa. It should be further mentioned that when we solve the Vlasov and Poisson's equations as an initial valve problem, here via $f_{0,-}+f_{0,+}$, it is possible to obtain the solutions with negative or positive values of $\omega_{i}$, corresponding to the damped or growing waves, respectively.
This can be explicitly seen from the electrostatic potential associated with the wave number $k$ of the excitation as follows:
\begin{equation}
\phi_{1}(x,t)=\phi_{1}e^{i(kx-\omega_{r} t)}e^{\omega_{i}t}, \label{phi1}
\end{equation}
where a solution with negative $\omega_{i}$ displays a damped wave, while the solution with positive one corresponds to an unstable mode.

When the damping or growth is weak we can expand the velocity integral in Eq. (\ref{DR}) around $\omega=\omega_{i}$ to find the zeros of $D(k,\omega)$. The dielectric function $D(k,\omega)$ is in general a complex function and thus the dispersion relation can be written as follows:
\begin{equation}
D(k,\omega_{r}+i\omega_{i})=D_{r}(k,\omega_{r}+i\omega_{i})+i D_{i}(k,\omega_{r}+i\omega_{i})=0, \label{Dtot}
\end{equation}
where $D_{r}$ and $D_{i}$ are the real and imaginary parts of the dielectric function. Since we want to consider the weakly damped or growing waves, i.e., $\omega_{i}\ll \omega_{r}$, the Eq. (\ref{Dtot}) can be Taylor expanded in the small quantity $\omega_{i}$ as follows:
\begin{equation}
D_{r}(k,\omega_{r})+i\omega_{i} \frac{\partial D_{r}(k,\omega_{r})}{\partial \omega_{r}}+i[D_{i}(k,\omega_{r})+i\omega_{i} \frac{\partial D_{i}(k,\omega_{r})}{\partial \omega_{r}}], \label{Dexpand}
\end{equation}
where $D_{r}$ and $D_{i}$ read
\begin{equation}
D_{r}(k,\omega_{r})=1-\frac{4 \pi n_{0} e^2}{mk^2} P.V.\int \frac{\frac{\partial}{\partial u}(f_{0,-} (u)+f_{0,+}(u))}{u-\frac{\omega_{r}}{k}}\,\mathrm{d}u, \label{Dr}
\end{equation}
\begin{equation}
D_{i}(k,\omega_{r})=-\pi (\frac{4 \pi n_{0} e^2}{mk^2}) [\frac{\partial}{\partial u} (f_{0,-} (u)+f_{0,+}(u)) ]_{u=\frac{\omega_{r}}{k}}. \label{Di}
\end{equation}
Here, we have made the analytic continuation of the velocity integral of the Eq. (\ref{DR}) over $u$, along the real axis, which passes under the pole at $u=\frac{\omega}{k}$ with the constraint of weakly damped waves, where $P.V.\int$ denotes the Cauchy principal value.
With the assumption $\omega_{i}\ll\omega_{r}$, by balancing the real and imaginary parts of the Eq.(\ref{Dexpand}) and neglecting the terms of order $(\frac{\omega_{i}}{\omega_{r}})^2$, we find that $\omega_{r}$ and $\omega_{i}$ can be computed, respectively, from the relations
\begin{subequations}
\begin{eqnarray}
D_{r}(k,\omega_{r})=0,
\\
\omega_{i}=-\frac{D_{i}(k,\omega_{r})}{{\partial D_{r}(k,\omega_{r})}/{\partial \omega_{r}}}. \label{findomegai}
\end{eqnarray}
\end{subequations}

\section{Acoustic modes with nonextensive stationary state}

Now, we want to obtain the formalism and some features of the acoustic-like modes in a pair (electron-positron or pair-ion) plasma in the context of the Tsallis nonextensive statistics. For this purpose we assume that the stationary state of the plasma obeys the $q$-nonextensive distribution function, instead of a Maxwellian one, which merely describes a fully equilibrium plasma. The $q$-nonextensive distribution function of stationary state for species $\alpha$ in one-dimension is given by \citep{Maxwell ansatz,mean value constraint1,mean value constraint2,H-theorem}
\begin{equation}
f_{0\alpha}(u)=A_{\alpha,q} [1-(q-1) \frac{m_{\alpha}u^2}{2k_{B}T_{\alpha}}]^{\frac{1}{q-1}}, \label{fq}
\end{equation}
where $m_{\alpha}$ and $T_{\alpha}$ are, respectively, the mass and temperature of species $\alpha$ ($\alpha=e^{\pm} \  \mathrm{or} \ \mathrm{C}_{60}^{\pm} $) and $k_{B}$ is the standard Boltzmann constant. The normalization constant $A_{\alpha,q}$ can be written as
\begin{equation}
A_{\alpha,q}=L_{q}\sqrt{\frac{m_{\alpha}}{2\pi k_{B}T_{\alpha}}},
\end{equation}
where the dimensionless $q$-dependent coefficient $L_{q}$ reeds
\begin{subequations}
\begin{eqnarray}
L_{q}=\frac{\Gamma (\frac{1}{1-q})}{\Gamma (\frac{1}{1-q}-\frac{1}{2})}\sqrt{1-q}, \ \ \ \  \mathrm{for}\ \ -1<q\leq1
\\
L_{q}=(\frac{1+q}{2})\frac{\Gamma (\frac{1}{2}+\frac{1}{q-1})}{\Gamma (\frac{1}{q-1})}\sqrt{q-1}. \ \ \ \ \mathrm{for}\ \ q\geq1
\end{eqnarray}
\label{Lq}
\end{subequations}
One  may examine that for $q>1$ , the $q$-distribution function (\ref{fq}) exhibits a thermal cutoff, which limits the velocity of particles to the values $u<u_{max}$, where $u_{max}=\sqrt{\frac{2k_{B}T_{\alpha}}{m_{\alpha} (q-1)}}$. For these values of the parameter $q$ we have $ S_{q>1}(A+B)<S(A)+S(B) $ referred to the \emph{subextensivity}. This thermal cutoff is absent when $q<1$ , that is, the velocity of particles is unbounded for these values of the parameter $q$. In this case, we have $ S_{q<1}(A+B)>S(A)+S(B) $ referred to the \emph{superextensivity}. Moreover, the $q$-nonextensive distribution (\ref{fq}) is unnormalizable for the values of the $q<-1$. Furthermore, the parameter $q$ may be further restricted by the other physical requirements, such as finite total number of particles and consideration of the energy equipartition for contribution of the total mean energy of the system. Interestingly, in the extensive limit $q\rightarrow1$ where $ S(A+B)=S(A)+S(B) $, and by using the formula $lim_{\mid z \mid\rightarrow\infty}z^{-a}[\frac{\Gamma (a+z)}{\Gamma (z)}]=1$ \citep{Abramowitz}, the distribution function (\ref{fq}) reduces to the standard Maxwell-Boltzmann distribution $ f_{0\alpha}(u)=\sqrt{\frac{m_{\alpha}}{2\pi k_{B}T_{\alpha}}} e^{-\frac{m_{\alpha}u^2}{2k_{B}T_{\alpha}}} $.
In Fig. 1, we have depicted schematically the nonthermal behavior of the distribution function (\ref{fq}) for some values of the spectral index $q$ in which the velocity $u$ and the distribution function $f(u)$ have normalized by the standard thermal speed $v_{th}=\sqrt{\frac{2k_{B}T}{m}}$ and $\sqrt{\frac{m}{2 \pi k_{B}T}}$, respectively.
We can see that in the case of a superextensive distribution with $q<1$ [Fig. 1(a)], comparing with the Maxwellian limit (solid curve), there are more particles with the velocities faster than the thermal speed $v_{th}$. These are the so-called superthermal particles and we can see that the $q$-distribution with $q<1$ behave like the $\kappa$ distribution, the same as that introduced for the first time by Vasyliunas in 1968 to describe the space plasmas far from the thermal equilibrium \citep{Vasyliunas}. In fact, in a superthermal plasma modeled by a $\kappa$-like distribution (here, the cases in which $q<1$), the particles have distributed in a wider spectrum of the velocities, in comparison with a Maxwellian plasma. In other words, the low values of the spectral index $q$ correspond to a large fraction of superthermal particle populations in the plasma.
On the other hand, in the case of a subextensive distribution with $q>1$ [Fig. 1(b)], comparing with the Maxwellian limit (solid curve), there is a large fraction of particles with the velocities slower than the thermal speed $v_{th}$. Moreover, for these values of the parameter $q$, we can explicitly see the mentioned thermal cutoff which limits the velocity of particles.
In fact, the $q$-nonextensive distributions with $q>1$ are suitable for describing the systems containing a large number of low speed particles.

The phase velocity of the acoustic modes in a pair plasma lies between the thermal velocities of the pairs. Here, we assume that $T_{+}<T_{-}$ and therefore the phase velocity of the acoustic waves lies in the frequency band $v_{th,+}<v_{\phi}<v_{th,-}$, where $v_{\phi}=\frac{\omega_{r}}{k}$ and $v_{th,\pm}=(\frac{k_{B}T_{\pm}}{m})^{\frac{1}{2}}$, respectively, denote the phase velocity of the wave and thermal speed of the pairs.
It is to be noted that because of the symmetry involved in a pair plasma, the other case in which $T_{-}<T_{+}$ is physically identical to our assumption here. Moreover, it is reminded that because of the same dynamics of the species in a pure pair plasma, we do not make a considerable difference in temperatures of the pairs, but we assume that it is finite and small. As we mentioned earlier, we may postulate physically that this finite temperature-asymmetry in a pair plasma may arise from the typical experimental procedure in which the pair plasma is produced in the laboratory \citep{ep production1,ep production2,ep production3,ep production4}.

With $v_{th,+}<v_{\phi}<v_{th,-}$, the Cauchy principal value of Eq. (\ref{Dr}) for the terms that are involving $f_{0,-}$ and $f_{0,+}$ may be evaluated by an expanding in $u$ as follows:
\begin{subequations}
\begin{eqnarray}
\int^{+u_{max}}_{-u_{max}} \frac{\frac{\partial }{\partial u} f_{0,-}(u) }{u-\frac{\omega_{r}}{k}}\,\mathrm{d}u=\int^{+u_{max}}_{-u_{max}} \frac{\partial f_{0,-}(u)}{\partial u}(\frac{1}{u}+\frac{1}{u^2}\frac{\omega_{r}}{k}+\frac{1}{u^3}\frac{\omega_{r}^2}{k^2}+...)\,\mathrm{d}u,
\\
\int^{+u_{max}}_{-u_{max}} \frac{\frac{\partial }{\partial u} f_{0,+}(u) }{u-\frac{\omega_{r}}{k}}\,\mathrm{d}u=-\frac{k}{\omega_{r}} \int^{+u_{max}}_{-u_{max}} \frac{\partial f_{0,+}(u)}{\partial u}(1+\frac{k}{\omega_{r}}u+\frac{k^2}{\omega_{r}^2}u^2+\frac{k^3}{\omega_{r}^3}u^3+...)\,\mathrm{d}u.
\end{eqnarray} \label{expand}
\end{subequations}
Here, in order to include both cases $q<1$ (superextensivity) and $q>1$ (subextensivity), we have denoted the integration limits in Eq. (\ref{expand}) by $\pm u_{max}$. In fact, as discussed earlier, the integration limits are unbounded, i.e., $\pm u_{max}=\pm \infty$, when $q<1$, and they are given by the $q$-dependent thermal cutoff $\pm u_{max}=\pm \sqrt{\frac{2k_{B}T_{\alpha}}{m_{\alpha}(q-1)}}$ when $q>1$.

With the $q$-nonextensive distribution given in Eq. (\ref{fq}), noting that $f_{0\alpha}(u)$ is an even function with argument $u$ and $\frac{\partial f_{0\alpha}}{\partial u}$ is an odd function, we may calculate the real part of the dielectric function in Eq. (\ref{Dr}) as follows:
\begin{equation}
D_{r}(k,\omega_{r})=1+\frac{4 \pi n_{0} e^2}{m k^2}\frac{1}{v_{th,-}^2}(\frac{1+q}{2})
-\frac{4 \pi n_{0} e^2}{m \omega_{r}^2}[1+3(\frac{2}{3q-1})\frac{k^2}{\omega_{r}^2}v_{th,+}^2].
\label{drfinal}
\end{equation}
The integrals in Eq. (\ref{expand}) are computed by parts and there, we have calculated the average values of $u^2$ as follows:
\begin{equation}
<u^2>= \int^{+u_{max}}_{-u_{max}} u^2 f_{\alpha0}(u) \,\mathrm{d}u=\frac{2}{3q-1}\frac{k_{B}T_{\alpha}}{m_{\alpha}}, \label{equi}
\end{equation}
which requires that the parameter $q$ must restrict to the values of $q>\frac{1}{3}$. Note that for the values of $q$ equal or lower than the critical value $q_{c}=\frac{1}{3}$, the mean value of $u^2$ diverges. Therefore, we see that the parameter $q$ for the case $q<1$ is further restricted to the values $\frac{1}{3} <q<1$, in order that the physical requirement of energy equipartition is preserved.
We emphasize that our results here are valid both for the case $\frac{1}{3}<q<1$ where the value of $u_{max}$ is unbounded and also in the case $q>1$ in which $u_{max}$ is given by the thermal cutoff $u_{max}=\sqrt{\frac{2k_{B}T_{\alpha}}{m_{\alpha}(q-1)}}$. Note that in both cases the integrals in Eq. (\ref{expand}) are evaluated by limits that are symmetric across the origin. The interested reader may easily check the validity of Eqs. (\ref{drfinal}) and (\ref{equi}) for all allowed values of $q$. Furthermore, in the extensive limit $q\rightarrow1$, Eq.~(\ref{equi}) reduces to the familiar energy equipartition theorem for each degree of freedom in the BG statistics as $<\frac{1}{2}m_{\alpha}u^2>=\frac{1}{2}k_{B}T_{\alpha}$.

It is to be noted that the $q$ distribution given in Eq. (\ref{fq}) describes the stationary state of the species $\alpha$ in the framework of the Tsallis nonextensive formalism. The value of the spectral index $q$ is a measure that determines the slope of the energy spectrum of the nonthermal particles and measures the deviation from the standard thermal distribution (which is recovered at the limit $q\rightarrow1$). The value of the spectral index $q$ is determined as a result of long-range interactions and correlations of the whole system. Therefore, a distinction between the pairs in $q$ can be or not, depend on the physics of the system under consideration. Here, following El-Tantawy \emph{et al.} \citep{El-Tantawy}, we make no distinction between the pairs in $q$.

The solution of the equation $D_{r}(k,\omega_{r})=0$ may yield the dispersion relation for the acoustic modes in a nonextensive pair plasma as follows:
\begin{equation}
\omega_{r}^2=k^2 c_{s}^2 [\frac{1}{(k \lambda_{D})^2(1+\frac{1}{\sigma})+(\frac{1+q}{2})}+3(\frac{2}{3q-1})\sigma ],
\label{omegr}
\end{equation}
where we have defined the sound-speed of the acoustic-like modes as $c_{s}={(\frac{k_{B} T_{-}}{m})}^{\frac{1}{2}}$. Here, $\sigma=\frac{T_{+}}{T_{-}}$ is the fractional temperature of positive to negative species and $\lambda_{D}$ is the Debye screening length and is given in a charge-neutral pair plasma by
\begin{equation}
\lambda_{D}^{-2}=\frac{4 \pi n_{0} e^2}{k_{B}}(\frac{1}{T_{-}}+\frac{1}{T_{+}}). \label{lambdaD}
\end{equation}
By definition of the the natural oscillation frequency in a charge-neutral pair plasma as $\omega_{p}=(\frac{8 \pi n_{0} e^2}{m})^{\frac{1}{2}}$ \citep{Saberian}, it is convenient to rewrite the linear dispersion relation for the later references as follows:
\begin{equation}
(\frac{\omega_{r}}{\omega_{p}})^2=(k \lambda_{D})^2 [\frac{\frac{1}{2}(1+\frac{1}{\sigma})}{(k \lambda_{D})^2(1+\frac{1}{\sigma})+(\frac{1+q}{2})}+3(\frac{1}{3q-1})(1+\sigma)].
\label{omegr2}
\end{equation}
On the other hand, by using the Eq. (\ref{Di}) and applying the $q$-nonextensive distribution function (\ref{fq}), it is straightforward to obtain the imaginary part of the dielectric function as follows:
\begin{equation}
D_{i}(k,\omega_{r})=L_{q} \frac{\sqrt{\pi}}{k^3 \lambda_{D}^3 (1+\frac{1}{\sigma})^{\frac{3}{2}}}\frac{\omega_{r}}{\omega_{p}}\{[1-(q-1)\frac{\omega_{r}^2}{k^2 \lambda_{D}^2 \omega_{p}^2 (1+\frac{1}{\sigma})}]^{\frac{2-q}{q-1}}+ \frac{1}{\sigma^{\frac{3}{2}}}[1-(q-1)\frac{\omega_{r}^2}{k^2\lambda_{D}^2 \omega_{p}^2 (1+\sigma)}]^{\frac{2-q}{q-1}}\}. \label{difinal}
\end{equation}
By $D_{r}(k,\omega_{r})$ and $D_{i}(k,\omega_{r})$ given in Eqs. (\ref{drfinal}) and (\ref{difinal}), we may obtain the explicit solution of the imaginary part of the frequency by using the relation (\ref{findomegai}), noting that both $k\lambda_{D}$ and $\frac{\omega_{i}}{\omega_{r}}$ are assumed small. The result is as follows:
\begin{eqnarray}
\omega_{i}=-\sqrt{\frac{\pi}{8}} \omega_{r} L_{q} (\frac{1}{(k \lambda_{D})^2(1+\frac{1}{\sigma})+(\frac{1+q}{2})}+3(\frac{2}{3q-1})\sigma)^{\frac{3}{2}} \times  \nonumber \\
\{ [1-(q-1)( \frac{\frac{1}{2}}{(k \lambda_{D})^2(1+\frac{1}{\sigma})+(\frac{1+q}{2})}+\frac{3}{2}(\frac{2}{3q-1})\sigma  ) ]^{\frac{2-q}{q-1}}+
\nonumber \\
\frac{1}{\sigma^{\frac{3}{2}}}
[1-(q-1) ( \frac{\frac{1}{2\sigma}}{(k \lambda_{D})^2(1+\frac{1}{\sigma})+(\frac{1+q}{2})}+\frac{3}{2}(\frac{2}{3q-1}))  ]^{\frac{2-q}{q-1}} \},
\label{omegi}
\end{eqnarray}
where $L_{q}$ is that given in Eq. (19).

Note that in deriving the solutions (\ref{omegr}) and (\ref{omegi}) for the acoustic-like modes in a pair plasma, we have considered the condition $k\lambda_{D}\ll1$ which indicates the regions with weak damping or growth (long wavelength limit).
Moreover, the values of the parameter $\sigma$ (the fractional temperature of the species) must be considered at the vicinity of unit, in order that a suitable compatibility with the physical circumstances is preserved.

In the extensive limit $q\rightarrow1$, our results reduce to the solutions for the acoustic-like modes in a Maxwellian pair plasma as follows:
\begin{equation}
\omega_{r}^2=k^2 c_{s}^2 [\frac{1}{k^2 \lambda_{D}^2(1+\frac{1}{\sigma})+1}+3\sigma ] \label{Maxwellian0}
\end{equation}

\begin{equation}
\frac{\omega_{i}}{\omega_{r}}=-\sqrt{\frac{\pi}{8}} (\frac{1}{k^2 \lambda_{D}^2 (1+\frac{1}{\sigma})+1}+3\sigma)^{\frac{3}{2}} \{ e^{-(\frac{\frac{1}{2}}{k^2 \lambda_{D}^2 (1+\frac{1}{\sigma})+1}+\frac{3}{2}\sigma )}
+\frac{1}{\sigma^{\frac{3}{2}}}e^{- (\frac{\frac{1}{2\sigma}}{k^2 \lambda_{D}^2(1+\frac{1}{\sigma})+1}+\frac{3}{2} )}  \} \label{Maxwellian}
\end{equation}
Note that in the extensive limit, the acoustic waves have only the (Landau) damping and no growth, because of the negative value of the imaginary part of the frequency, provided by Eq. (\ref{Maxwellian}). Furthermore, in the symmetric case $\sigma\rightarrow1$, the dispersion relation of the acoustic waves in pair or pair-ion plasmas given in Eq.(\ref{Maxwellian0}), reduce to Eq.(12) of Ref. \citep{Kaladze}.
One basic feature of our work is the inclusion of the nonextensivity of the system, which is essentially as a result of the long-range Coulombic interactions of the charge particles in the plasma. The nonextensivity of the system is determined by the spectral index $q$ and may lead to positive or negative $\omega_{i}$ in Eq. (\ref{omegi}). Therefore, depending on the nonextensivity of the plasma, both the damped and growing acoustic modes may be happened in a pair plasma.

\section{Discussion}

\subsection{Dispersion relation}

The solutions (\ref{omegr2}) and (\ref{omegi}) describe the acoustic-like modes in a nonextensive electron-positron plasma or pair-ion plasma at the limit of long wavelengths confirmed by $k\lambda_{D}\ll1$. In {Fig. (2a)}\label{fig2} we have plotted the dispersion relation of acoustic modes for some values of the nonextensivity index $q$. In the represented graph, the solid curve corresponds to the extensive limit $q=1$ and the other ones show the deviations from the Maxwellian limit.

It is seen that for a given wavelength, the phase velocity of the acoustic modes increases with decreasing the value of $q$. The physical description can be discussed in the context of the nonextensive statistics as follows. As mentioned earlier, the $q$-distribution function with $q<1$, comparing with the Maxwellian one ($q=1$), indicates the systems with more superthermal particles, i.e., particles with the speed faster than the thermal speed $v_{th}=\sqrt{\frac{2k_{B}T}{m}}$ (superextensivity). On the other hand, the $q$-distribution with $q>1$ is suitable to describe systems containing a large number of low-speed particles (subextensivity). However, because of the long-range nature of Coulombic interactions in plasma environments and the presence of many superthermal particles in such systems, confirmed by many astrophysical measurements \citep{super1,super2,super3,super4}, a $q$-distribution with $q<1$ is strongly suggested for the real plasma systems or superthermal plasmas. It is obvious that in a plasma with more superthermal particles ($q<1$), the phase velocity of the acoustic-like modes should be larger than the case with lack of superthemal particles ($q>1$), in agreement with our results here.

In addition, we have illustrated the temperature-asymmetry effect, via $\sigma$, on the dispersion relation of acoustic modes in a pair plasma as shown in {Fig. 2(b)}\label{fig2}. There, the solid curve indicates the case in which the whole plasma is in a common thermal state with $T_{-}=T_{+}$, signifies a temperature-symmetric pair plasma, and the other curves show deviations from this symmetric case. We see that the temperature-asymmetry reduces the phase velocity of the acoustic modes in a pair plasma. However, our kinetic model confirms that the acoustic-like modes are possible in both symmetric and asymmetric pair plasmas, depart from a small shift in phase velocity.

It is reminded that in this work we have specialized our study to the low-frequency band in which $v_{th,+}<v_{\phi}<v_{th,-}$ . Then, the Cauchy principal value of Eq. (\ref{Dr}) is evaluated by an expanding in velocity in the form of Eq. (\ref{expand}). So, our calculations in this frequency band may lead to the acoustic modes and not to the Langmuir waves. On the other hand, considering a high frequency band in which the phase velocity of the wave is much larger than the thermal velocity of the particles ($v_{\phi}>>v_{th}$) may lead to the Langmuir-type waves, as studied in Ref. \citep{Saberian}. There, the dispersion relation for the Langmuir waves is given by
\begin{equation}
\omega_{r}^2=\omega_{p}^2[1+3(k\lambda_{D})^2\frac{2}{3q-1}].
\label{omegr-Lanqmuir}
\end{equation}
However, for comparison of the acoustic modes and the Langmuir waves in a pair plasma with $T_{+}=T_{-}$, we have depicted both of the acoustic and Langmuir branches in {Fig. 3}\label{fig3}. From this graph, we see explicitly that the acoustic waves belong to a low frequency band which tends to zero at the limit $k\rightarrow0$, while the Langmuir waves occur at high frequencies above $\omega_{p}$.

On the other hand, the experimental data presented by Oohara \emph{et. al} \citep{Oohara} confirm the possibility of the acoustic-like modes in a pair plasma which is compatible with our results here.

\subsection{Landau damping and unstable modes}

In {Fig. 4}\label{fig4} we have plotted the ratio $\omega_{i}/ \omega_{r}$ with respect to the nonextensivity index $q$ for all allowed values of $q<1$ (referred to superextensivity) at the limit of long wavelengths (supported by, e.g., $k\lambda_{D}=0.1$). It is seen that both of the damped ($\omega_{i}<0$) and growing ($\omega_{i}>0$) acoustic-like modes are predicted in a nonextensive pair plasma with $q<1$. Our numerical analysis shows that in the $q$-region $0.34\la q \la0.6$ the acoustic modes are unstable, due to the fact that $\omega$'s have positive imaginary parts and then the associated modes will grow in time (Eq. (\ref{phi1}) is reminded). The mechanism which leads to this instability may explain as follows. As we expressed earlier, the $q$-nonextensive distribution with $q<1$ describes a system with a large number of superthermal particles. So, our solution for the Vlasov and Poisson's equations with small values of $q<1$ indicates an evolution which has started from a stationary state with a large portion of superthermal particles. The acoustic-like waves may gain energy from these superthermal particles and results in growing waves in time. In other words, this instability arises from a stationary state which describes a superthermal plasma and, in fact, we have obtained a solution for acoustic-like modes in which the stationary sate of the plasma has started from a non-equilibrium distribution.
However, our results have the flexibility to reduce to the equilibrium solutions in the limiting case of $q\rightarrow1$ indicates a Maxwellian distribution.
Furthermore, the acoustic-like modes have Landau damping in the $q$-region $0.6\la q \la 0.71$ because $\omega$'s have negative imaginary parts in these degrees of the nonextensivity (see Fig. 4). The Landau damping is a resonance phenomena between the plasma particles and the wave, for the particles moving with nearly the phase velocity of the wave \citep{Landau paper,Krall}. Noting that the $q$-distribution is a decreasing function with $u$, there are more particles moving slightly slower than the wave than the particles moving slightly faster than the wave; if the slower particles are accelerated by the wave, this must reduce the energy of the wave, and the wave damps.

It is to be noted that our analysis shows that after $q=0.71$, the curve in Fig. 4 rises to positive values for a small interval of $q$ and then it returns to the negative values. The fluctuation of $\omega_{i}$ between the positive and negative values continues increasingly until to the limiting case at $q\rightarrow1$. In fact, the curve in Fig. 4 don't show a smooth behavior for the values $0.71<q<1$ and the analysis break down, until to the extensive limit at $q\rightarrow1$, where our solutions reduce smoothly to that of a Maxwellian pair plasma given in Eqs. (\ref{Maxwellian0}) and (\ref{Maxwellian}). This unsmooth behaviour is because of the existence of the terms $\Gamma(\frac{1}{1-q})$ and $\Gamma(\frac{1}{1-q}-\frac{1}{2})$ in our formalism supported by $L_{q}$. Indeed, this behavior is a mathematical consequence and there is not a physical justification for it. So, we have analyzed the problem in a well-defined interval of $q$, i.e, $1/3<q<0.71$, as shown in Fig. 4.

We can also investigate the resonance between the plasma particles and the acoustic modes for the values of $q>1$ (referred to subextensivity). In {Fig. 5}\label{fig5}, the ratio $\omega_{i}/ \omega_{r}$ with respect to nonextensivity index $q$ is plotted for the values of $q>1$ at a typical long wavelength ($k\lambda_{D}=0.1$). From this graph, it is seen that the acoustic-like modes have only (Landau) damping and no growth for these degrees of the nonextensivity. Furthermore, the damping rate is relatively weak in these $q$-regions, in comparison with the case of a superthermal plasma ($q<1$).
The reason is that the number of particles participating in the resonance with the wave is small for a stationary state with $q>1$. Strictly speaking, the slope of the velocity $q$-distribution function $f_{0\alpha}(u)$ given in Eq. (\ref{fq}) increases with $q$ and there is even a thermal cutoff in the case of $q>1$ [see Fig. 1(b)]. This corresponds to a weaker resonance with the wave, in comparison with the case $q<1$.

Our analysis reveals that the acoustic-like modes are unstable in the $q$-region $0.34\la q \la0.6$ (high superthermal $q$-region) yet, they are heavily damped in the $q$-region $0.6\la q \la 0.71$ (less superthermal $q$-region) and finally, they are relatively weakly damped for the values of $q>1$ (subextensive region).
In {Fig. 6}\label{fig5}, the damping and growing rates with respect to the wave number are plotted for some values of the nonextensive index $q$ for three cases of the heavily damped modes [Fig6(a)], weakly damped modes [Fig6(b)] and growing unstable modes [Fig6(c)]. We see that for the waves with longer wavelengths the rate of damping (or growth) becomes weaker. Moreover, our numerical analysis shows that in a pair plasma the acoustic-like modes have the maximum damping at the vicinity of $q=0.69$ [see Fig6(a)], and they have the maximum growth when the nonextensivity is at the vicinity of $q=0.55$ [see Fig6(c)].
In addition, we have included the Maxwellian limit ($q=1$) to the Fig. 6(b) which emphasizes that the acoustic-like modes in an equilibrium pair plasma are merely landau damped waves.

For completing our discussion, we have examined the temperature-asymmetry effect, controlled by $\sigma$, on the Landau damping of the acoustic-like modes in a pair plasma, as plotted in {Fig. 7}\label{fig7}. It is observed that the temperature-asymmetry in a pure pair plasma decreases the Landau damping. In other words, for a fixed value of $q$ and at a given wavelength, the Landau damping of the acoustic waves is maximum when a  full symmetry in temperature of species is established, i.e, when $T_{-}=T_{+}$.

\section{Conclusions}

In this paper, we have studied the acoustic-like modes in a collisionless and magnetic-field-free pair plasma on the basis of the nonextensive statistics. We have thereby used a kinetic theory model by employing the Vlasov and Poisson's equations to obtain the response dielectric function of the pair plasma for the electrostatic waves. By using the dielectric function, we have investigated the acoustic-like modes whose phase speed lies between the thermal velocities of the species.
The resultant dispersion relation in our study is compatible with the acoustic branch of the experimental data presented by Oohara \emph{et. al} \citep{Oohara}, in which the electrostatic waves have been examined in a pure pair-ion plasma.
It has been shown that by decreasing the nonextensivity index $q$ the phase velocity of the acoustic modes increases, indicating to a plasma with a great deal of superthermal particles.
Our kinetic model confirms the possibility of the acoustic modes in the case of a temperature-asymmetric and also symmetric pair plasma. However, it is found that the temperature-asymmetry in a pair plasma reduces the phase velocity of the acoustic modes.
Furthermore, depending on the degree of nonextensivity of the plasma, both the damped and unstable acoustic modes are predicted in a collisionless pair plasma, arise from a resonance phenomena between the wave and nonthermal particles of the plasma. In the case of a superthemal plasma confirmed by $q<1$ (superextensivity), the heavily damped and growing unstable modes are predicted, while in the case $q>1$ (subextensivity) the acoustic-like modes have only damping and no growth. The mechanism that leads to the damping is the same as presented by Landau \citep{Landau paper}, arises from a decreasing velocity distribution function, but the mechanism of instability lies in the heart of the nonextensive formalism. We have postulated that the concerned instability can be associated with the presence of superthermal particles (in the case $q<1$), in the sense that in the process of the resonance they can give energy to the wave and then results in growing waves in time.
This instability disappears in the case $q>1$, describing a plasma with plenty of the low-speed particles.
Additionally, the damping rate is relatively weak in the case $q>1$, in comparison with the case of a superthermal plasma ($q < 1$) with heavily damped modes. The reason is that the number of particles participating in the resonance with the wave is small for a stationary state with $q > 1$.
Moreover, our analysis indicates that the temperature-asymmetry in a pure pair plasma decreases the Landau damping of the acoustic-like modes.

We emphasize that in the present work, we have considered an inhomogeneous plasma in a nonequilibrium thermal state by considering the $q$-nonextensive distribution for stationary state of the plasma.
However, our solutions reduce to the ones for a homogeneous and equilibrium pair plasma at the extensive limit $q\rightarrow1$, in which the Boltzamnn-Gibbs statistics describe the plasma state.

It is hoped that the present study would be useful for explanation of the intriguing low-frequency modes in a pure pair plasma, which are out of the scope of the plasma fluid theory and the Boltzmann-Gibbs statistics.

\clearpage



\begin{figure}[H]
 \begin{center}
  \includegraphics{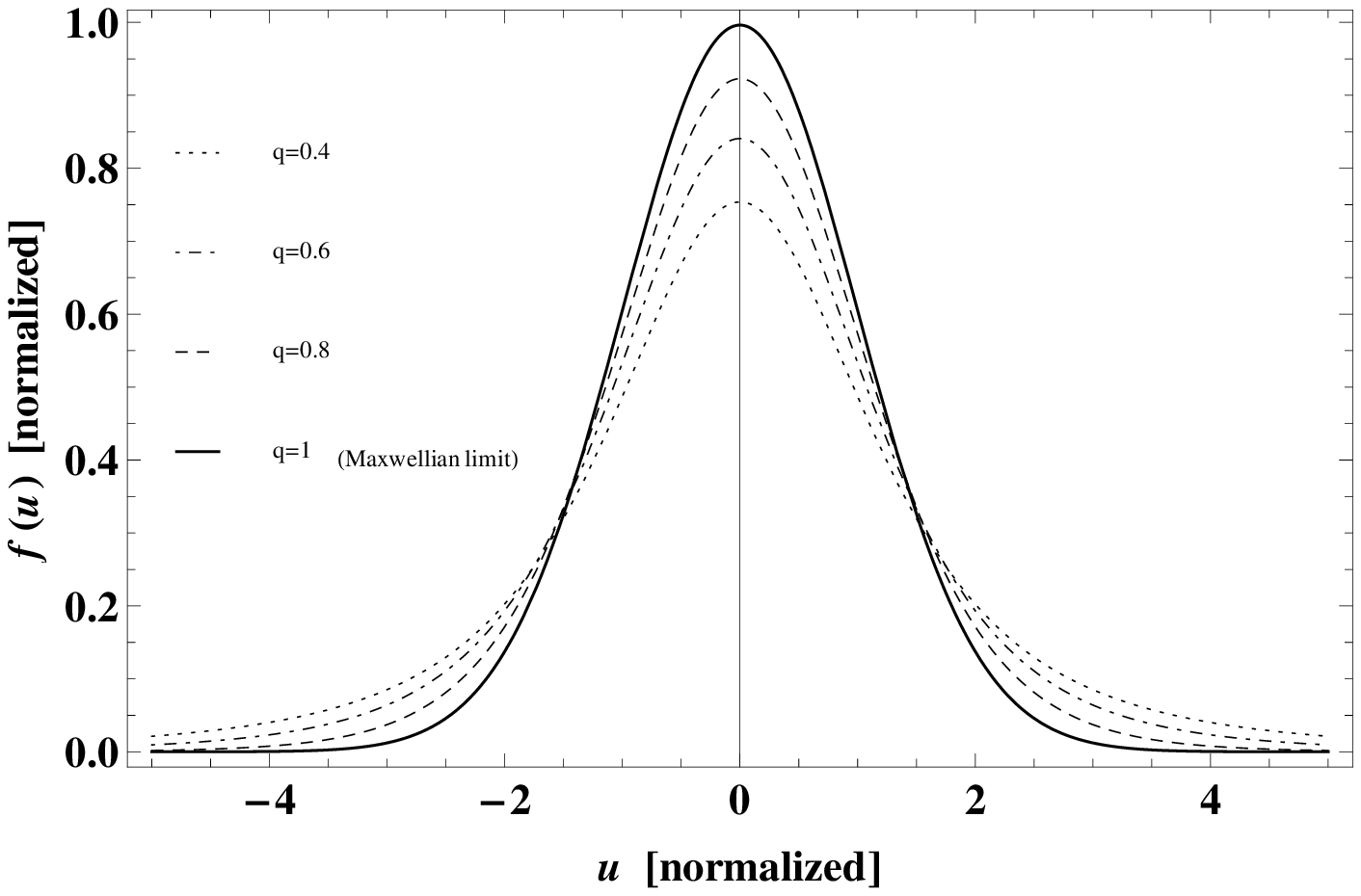}
  \label{fig1}
 \end{center}
\end{figure}

\begin{figure}[H]
 \begin{center}
  \includegraphics{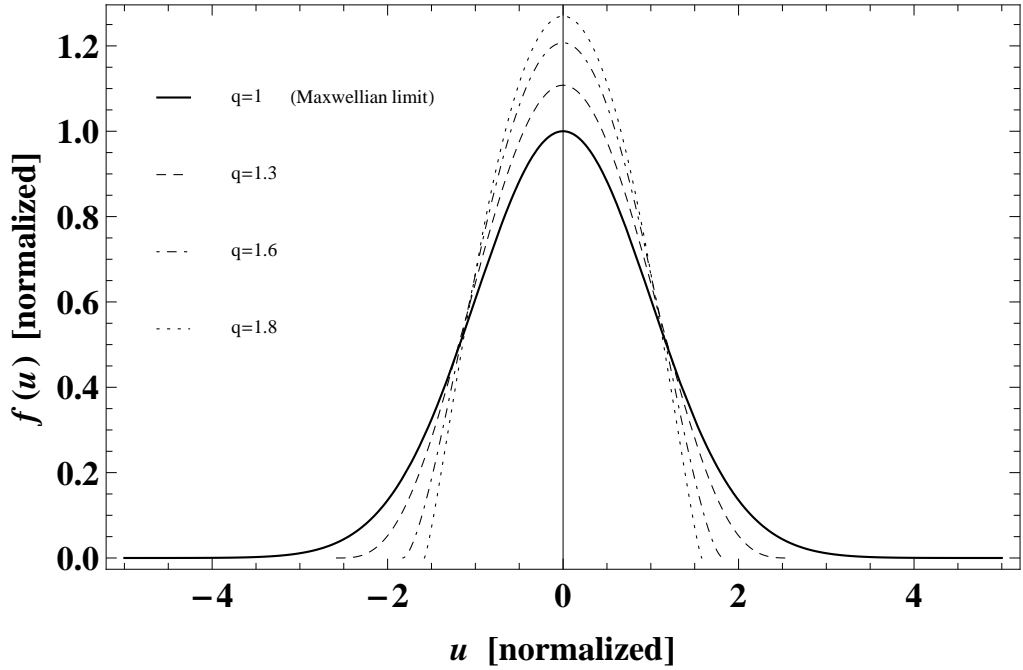}
\caption{The nonthermal behavior of the $q$-nonextensive distribution function and its comparison with the Maxwellian one (solid carve): (a) Superxtensive distribution with $q<1$ that behave alike the $\kappa$-distributions for superthermal plasmas. In this case, the particles have distributed in a wider spectrum of the velocities, in comparison with a Maxwellian distribution. (b) Subextensive distribution with $q>1$ which is suitable for describing the systems containing a large number of low-speed particles. In this case, there is a thermal cutoff which limits the velocity of particles.}
  \label{fig1}
 \end{center}
\end{figure}

\begin{figure}[H]
 \begin{center}
  \includegraphics{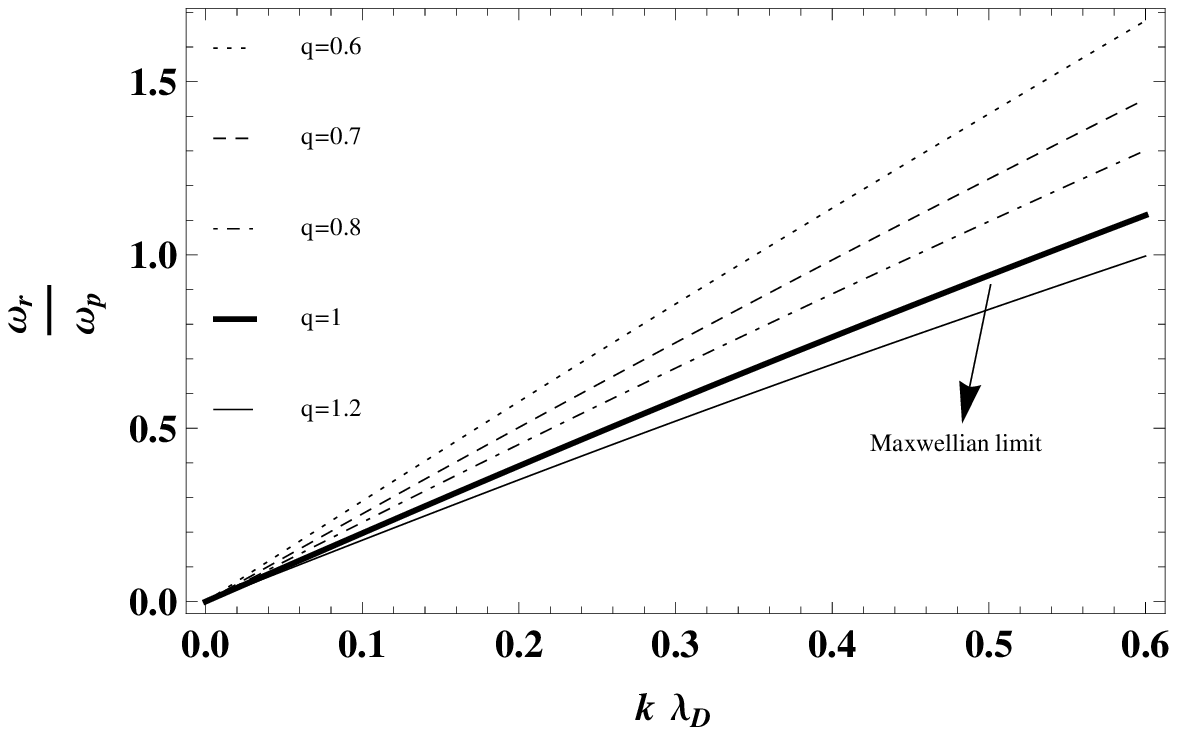}
  \label{fig2}
 \end{center}
\end{figure}

\begin{figure}[H]
 \begin{center}
  \includegraphics{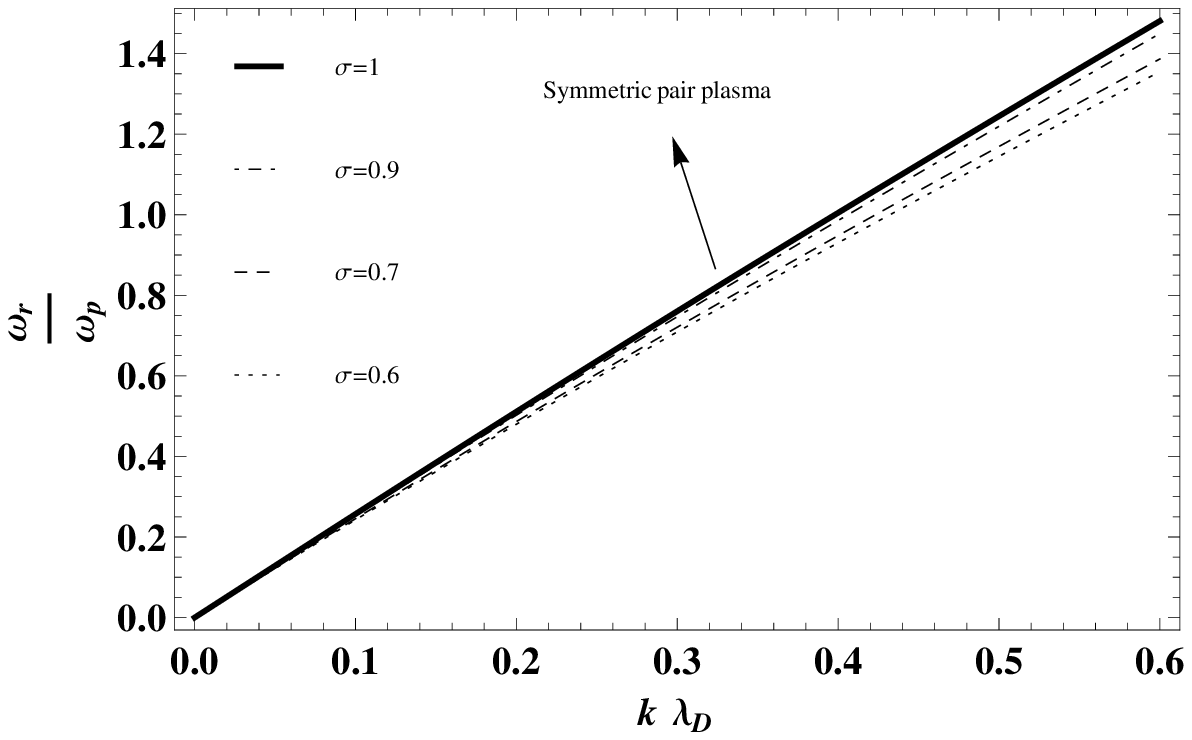}
\caption{The linear dispersion relation of acoustic-like modes in a pair plasma. (a) The nonextensivity effect on dispersion relation with $\sigma=0.9$, where the solid curve corresponds to the extensive limit ($q=1$) and the other ones show the deviations from a Maxwellian pair plasma. (b) The effect of temperature-asymmetry on dispersion relation with $q=0.7$, where the solid curve corresponds to a temperature-symmetric pair plasma.}
  \label{fig2}
 \end{center}
\end{figure}

\begin{figure}[H]
 \begin{center}
  \includegraphics{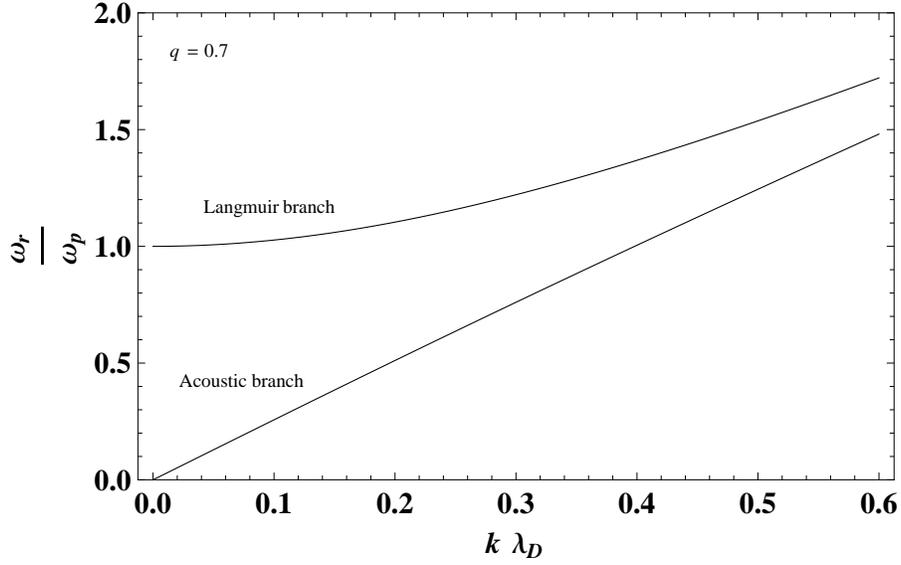}
\caption{The comparison of the acoustic-like modes and the Lanqmuir waves in a pair plasma with $T_{+}=T_{-}$. The acoustic waves belong to a low frequency band which tends to zero at the limit $k\rightarrow0$, while the Langmuir waves occur in high frequencies above $\omega_{p}$.}
  \label{fig3}
 \end{center}
\end{figure}

\begin{figure}[H]
 \begin{center}
  \includegraphics{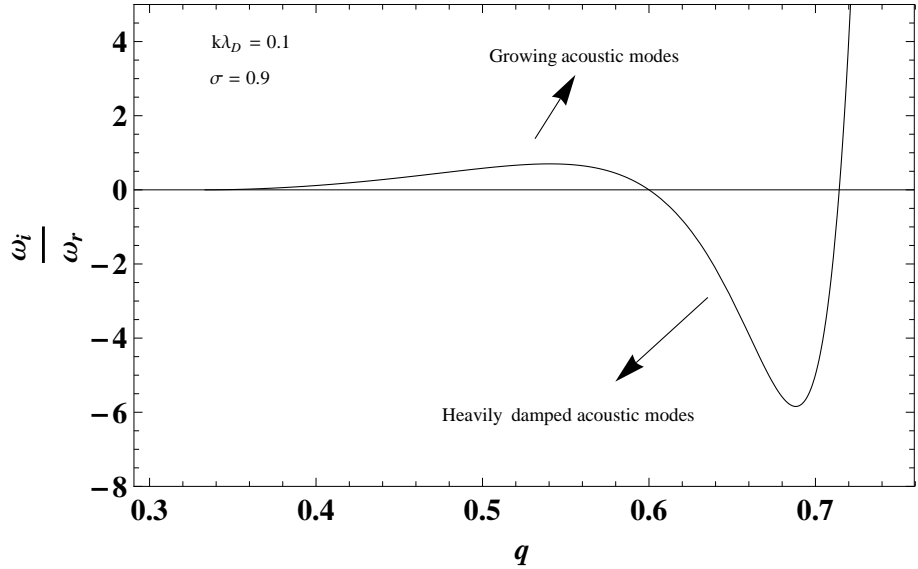}
  \caption{The imaginary part of the frequency with respect to the nonextensivity index for $q<1$, which shows the $q$-regions for the growing and heavily damped acoustic-like modes.}
  \label{fig4}
 \end{center}
\end{figure}

\begin{figure}[H]
 \begin{center}
  \includegraphics{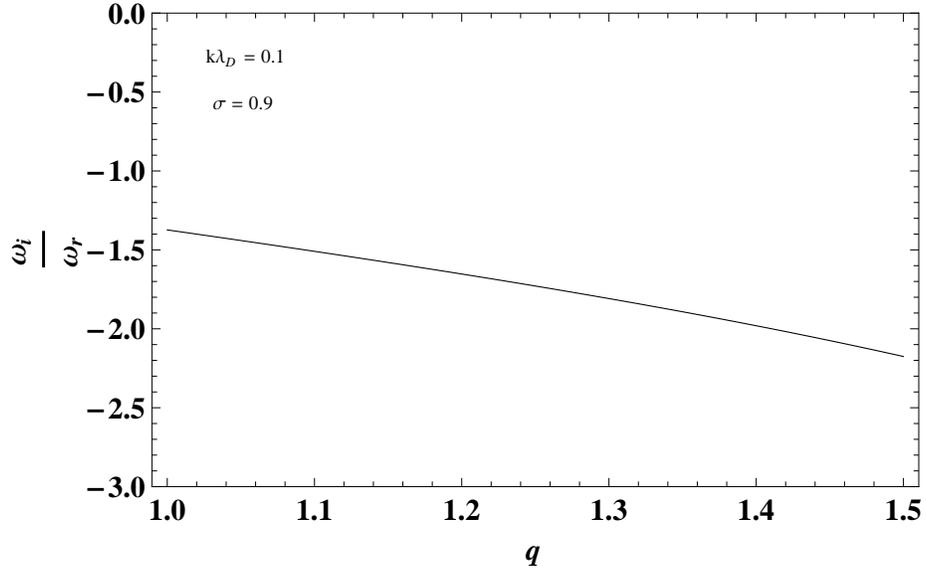}
  \caption{The imaginary part of the frequency with respect to the nonextensivity parameter for $q>1$. For this values of the nonextensivity index $q$, the acoustic-like modes have only damping and no growth.}
  \label{fig5}
 \end{center}
\end{figure}

\begin{figure}[H]
 \begin{center}
  \includegraphics{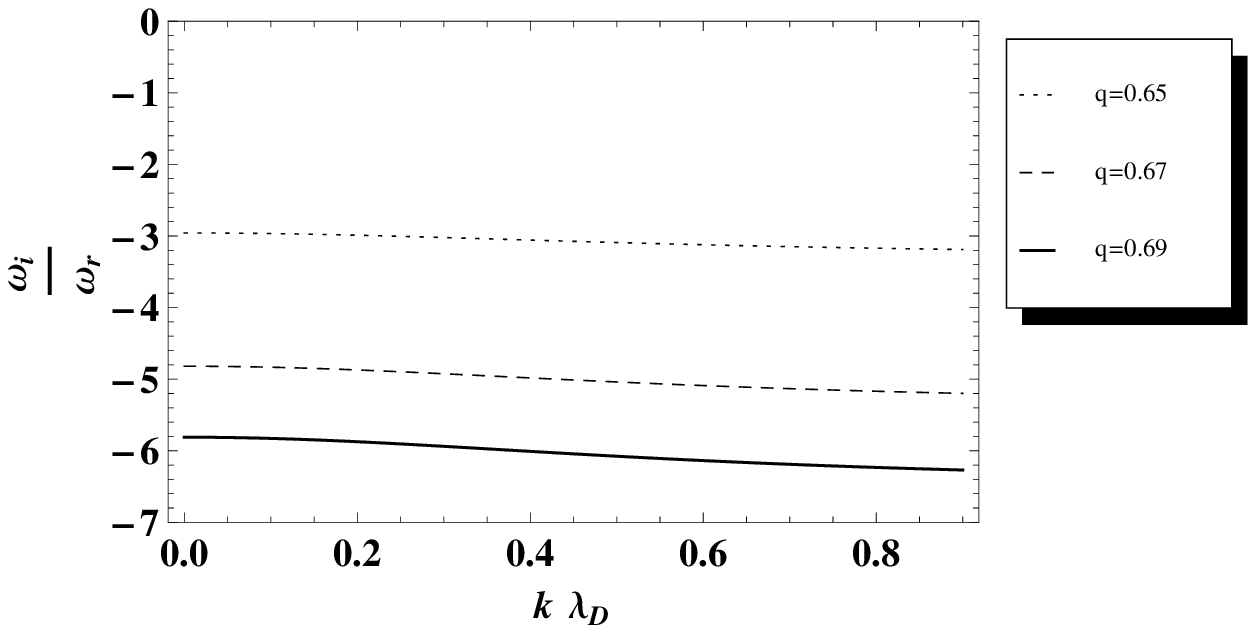}
  \label{fig6}
 \end{center}
\end{figure}

\begin{figure}[H]
 \begin{center}
  \includegraphics{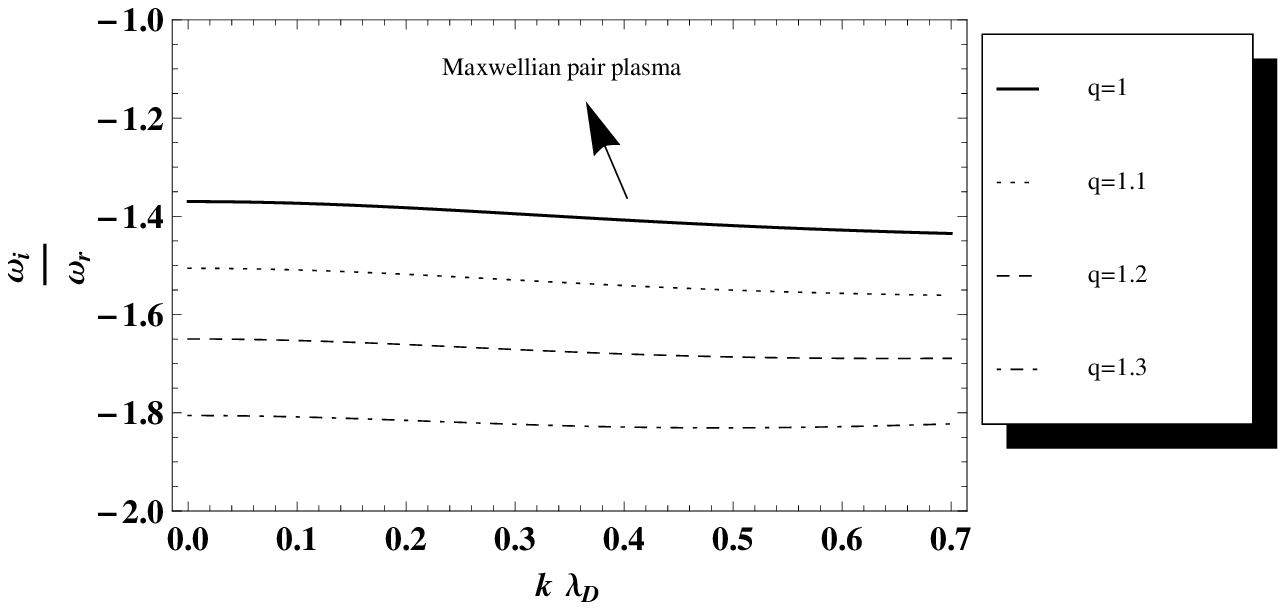}
  \label{fig6}
 \end{center}
\end{figure}

\begin{figure}[H]
 \begin{center}
  \includegraphics{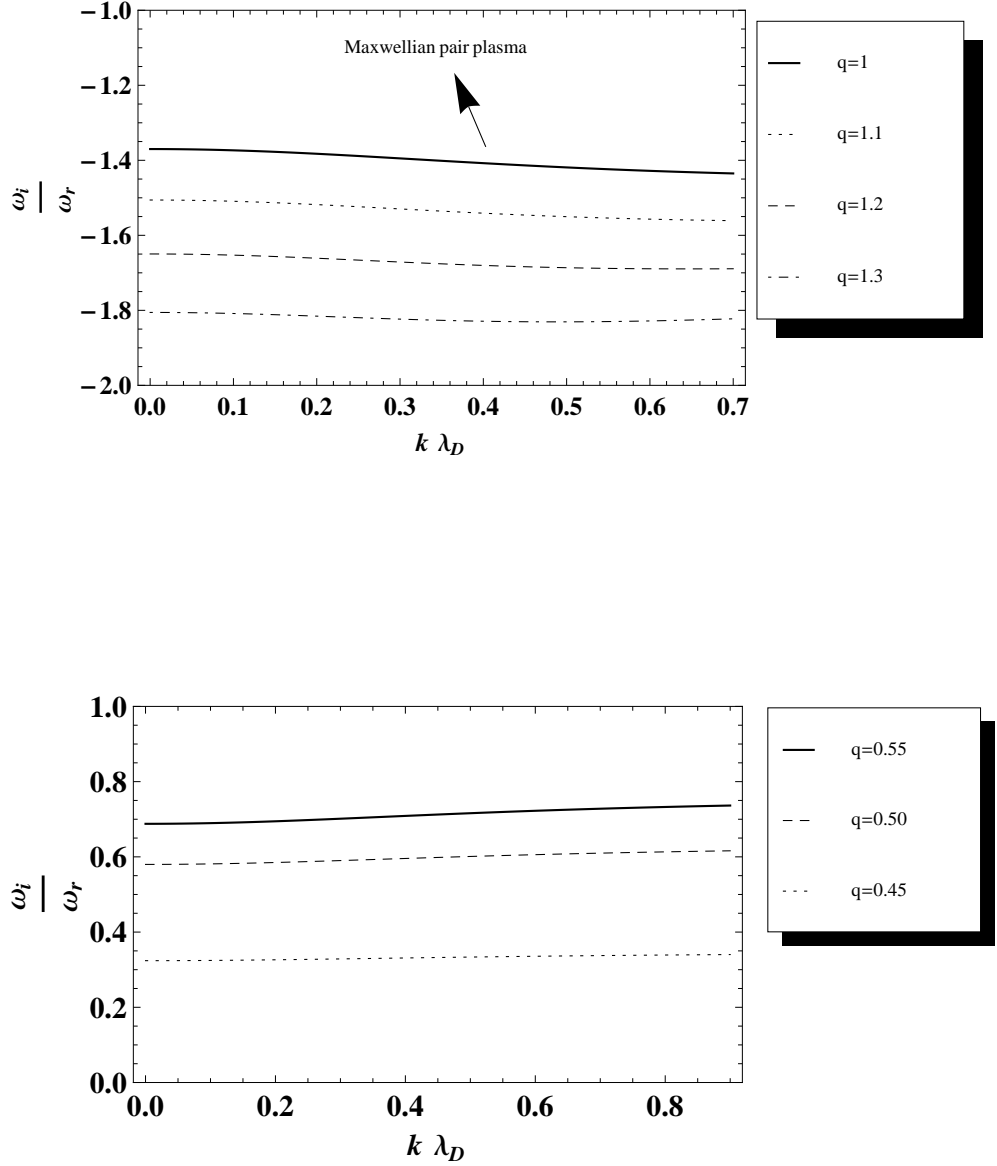}
\caption{The damping (growing) rate with respect to the wave number for (a) the heavily damped modes in the $q$-region $0.6\la q \la 0.71$, (b) the relatively weakly damped modes in the $q$-region $q>1$, and (c) the growing acoustic modes in the $q$-region $0.34\la q \la 0.6$, when $\sigma=0.9$. We have included the Maxwellian limit ($q=1$) to our results which emphasizes that the acoustic-like modes in an equilibrium pair plasma are merely the landau damped waves.}
  \label{fig6}
 \end{center}
\end{figure}

\begin{figure}[H]
 \begin{center}
  \includegraphics{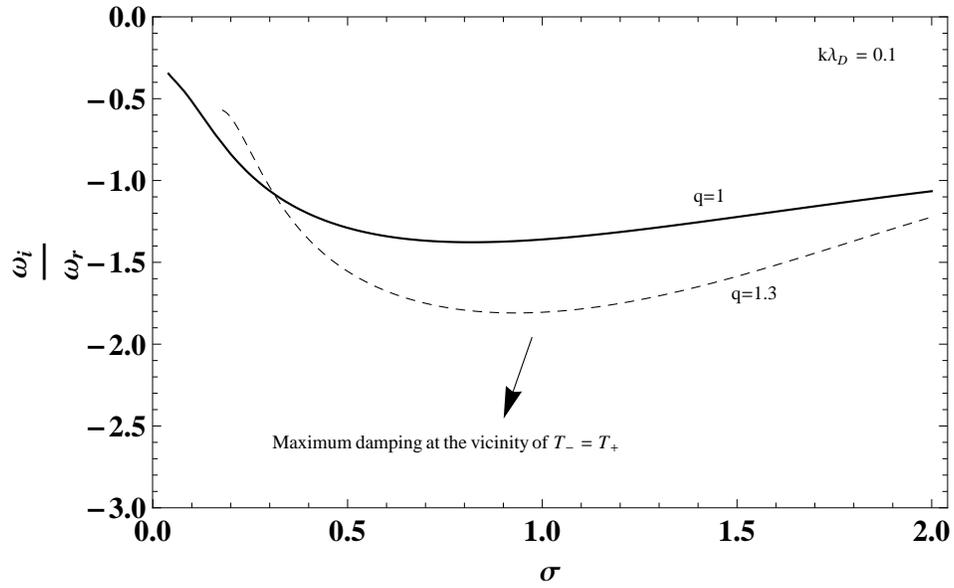}
\caption{The effect of temperature-asymmetry on Landau damping of the acoustic-like modes which indicates that the temperature-asymmetry in a pure pair plasma decreases the Landau damping rate.}
  \label{fig7}
 \end{center}
\end{figure}

\clearpage









\clearpage


\clearpage



\clearpage




\end{document}